\documentclass{camera}
\usepackage{graphicx,epsfig}  

\newcommand{\ipb}{\mbox{$\mathrm {pb}^{-1}$}}

\def\rs{\mbox{$\sqrt{s}$}}

\def\Journal#1#2#3#4{{#1} {\bf #2}, #3 (#4)}

\def\PRL{\em Phys. Rev. Lett.}

\begin{document}

%
\title{GAUGE BOSON SELF COUPLINGS \\AND FOUR FERMION FINAL STATES AT LEP}

%
\author{Paolo Azzurri}
%
\organization{Scuola Normale Superiore, 
Piazza dei Cavalieri 7, 56100 Pisa, Italy}

\maketitle

%
\abstract{
Four-fermion productions measured in the LEP2 data are reviewed.
The total and differential cross-section yields represent the 
first clear evidence for the existence of gauge boson self couplings, 
in support of the non-abelian SU(2)$\otimes$U(1) structure 
of the electroweak model, at the percent level.}

\section{Introduction}
\begin{figure}[htb]
  \centerline{\epsfig{file=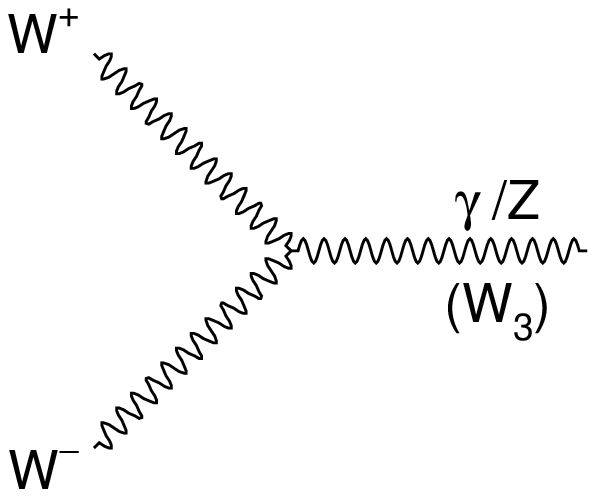,height=4cm}
              \epsfig{file=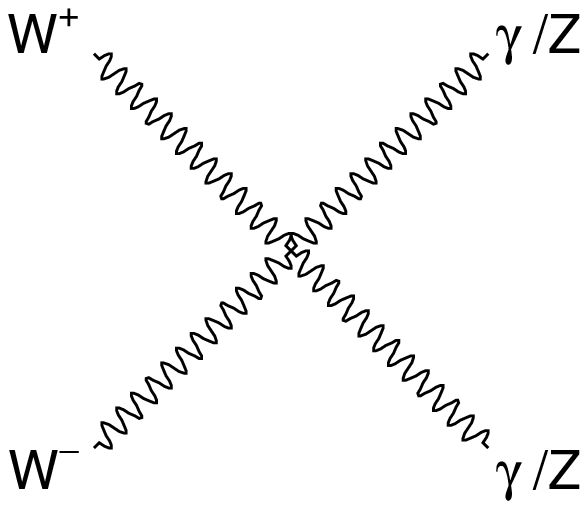,height=4cm}
              \epsfig{file=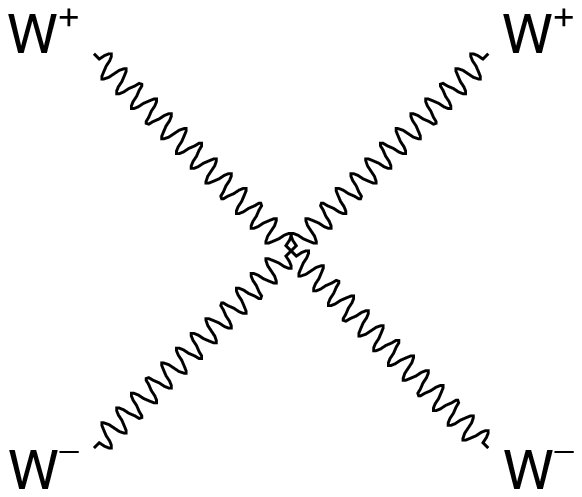,height=4cm}}
\caption{\sl Examples of triple and quartic gauge bosons 
self-couplings predicted by the SU(2) part of the electroweak model.}
\label{fig:bsf}
\end{figure}
A striking feature of the Standard Model (SM) 
of electroweak interactions~\cite{SM},
is the presence of gauge bosons self-couplings (GC) due to the 
non-abelian nature of the SU(2) group leading to 
three-line (TGC) and four-line (QGC) boson vertices 
shown in Fig.~\ref{fig:bsf}.

Gauge self couplings influence directly the 
pair-production of weak bosons, and in particular of W-bosons,
so that the LEP2 represents an ideal data sample to verify their 
structure, through the detection of the resulting
four-fermion final states.

\section{W-pair and Z-pair productions}
\begin{figure}[htb]
  \centerline{\epsfig{file=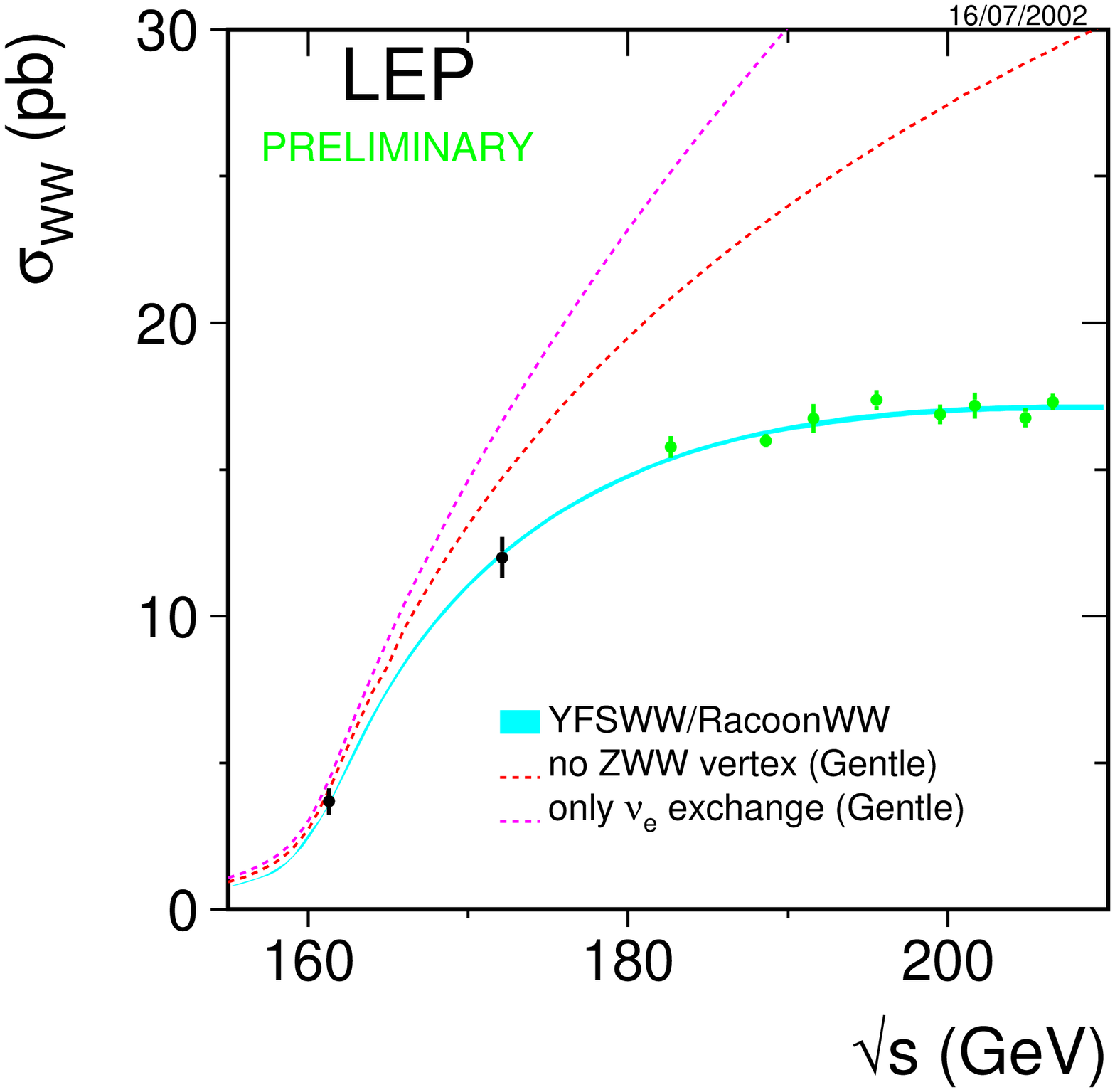,height=8cm}
              \epsfig{file=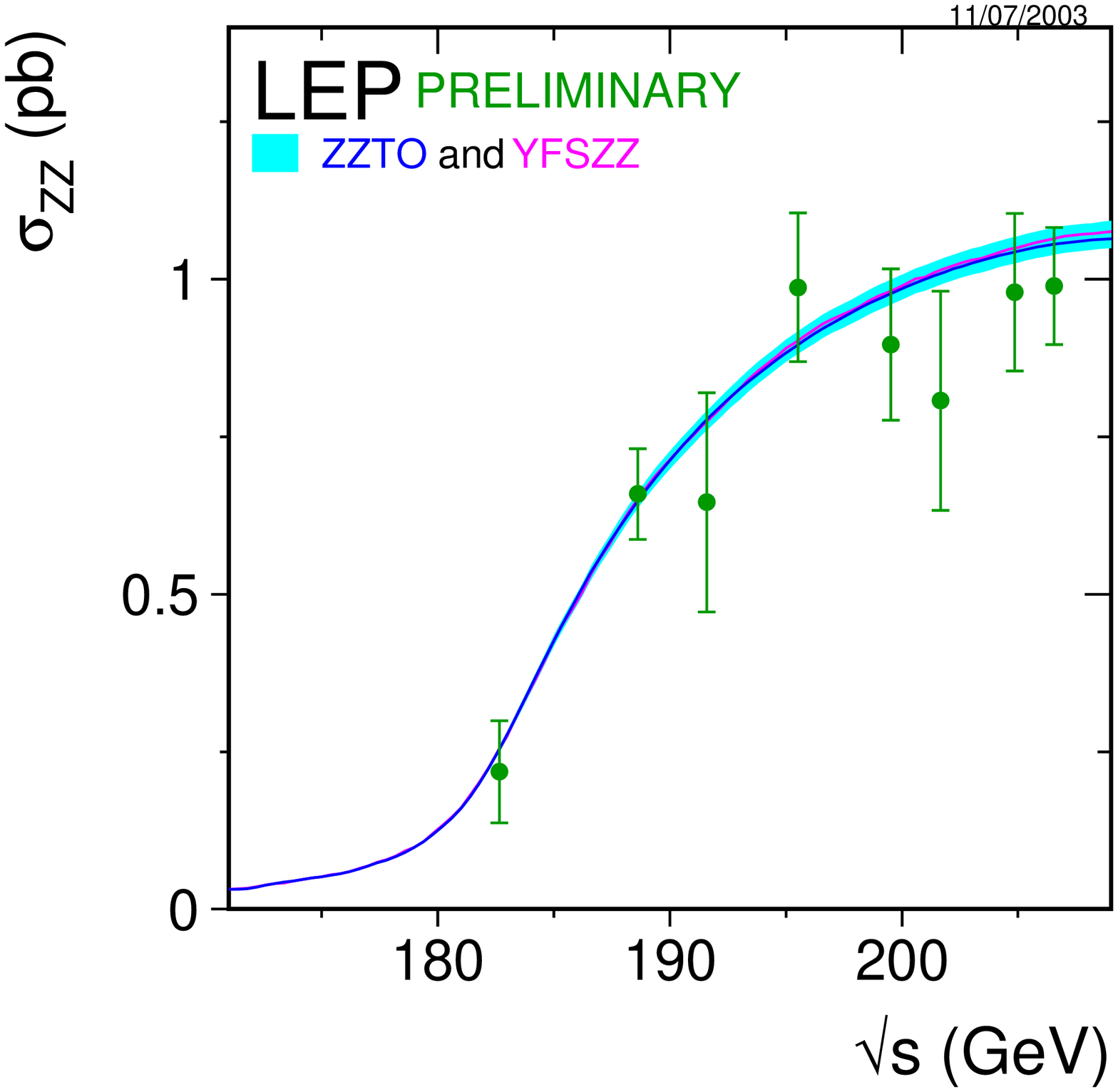,height=8cm}}
\caption{\sl Total W-pair and Z-pair cross-sections measured at LEP 
at $\rs$=161-207~GeV, compared to SM predictions. 
The two dashed curves show the predicted W-pair 
cross section in the absence of the WWZ and WW$\gamma$ couplings.}
\label{fig:xsec1}
\end{figure}

The LEP2 data sample of ~700$\ipb$ per experiment at $\rs$=161-207~GeV
allowed to collect about 10$^4$ W-pairs and 500 Z-pairs per experiment,
identifying these events in all their visible final states. 
The combined total cross sections, measured as a function of the 
centre-of-mass energy are shown in Fig.~\ref{fig:xsec1}.
The overall precision of the measured productions  
rates is at the level of 1\% for W-pairs and 5\% for Z-pairs, and
in agreement with the SM predictions.
As it can be seen in Fig.~\ref{fig:xsec1}, the W-pair cross 
section measurement represents alone a stunning proof of the 
presence of both the WWZ and WW$\gamma$ couplings dictated by the
electroweak SU(2)$\otimes$U(1) gauge structure.

The LEP2 W-pair sample has also proven valuable to provide 
the first direct measurement of all leptonic and hadronic
W decay branching ratios, resulting in a test of the lepton 
family universality of charged currents at the level of 1\%
($  g_\mu/g_{\rm e} = 1.000\pm 0.010,  
  g_\tau/g_{\rm e} =  1.026\pm 0.014, 
  g_\tau/g_{\mu} =  1.026\pm 0.014 $), 
and of the lepton-quark universality 
of charged currents at the level of 0.7\%
($g_{\rm q}/g_\ell= 1.010\pm 0.007$).

\section{Single W and single Z productions}
\begin{figure}[htb]
  \centerline{\epsfig{file=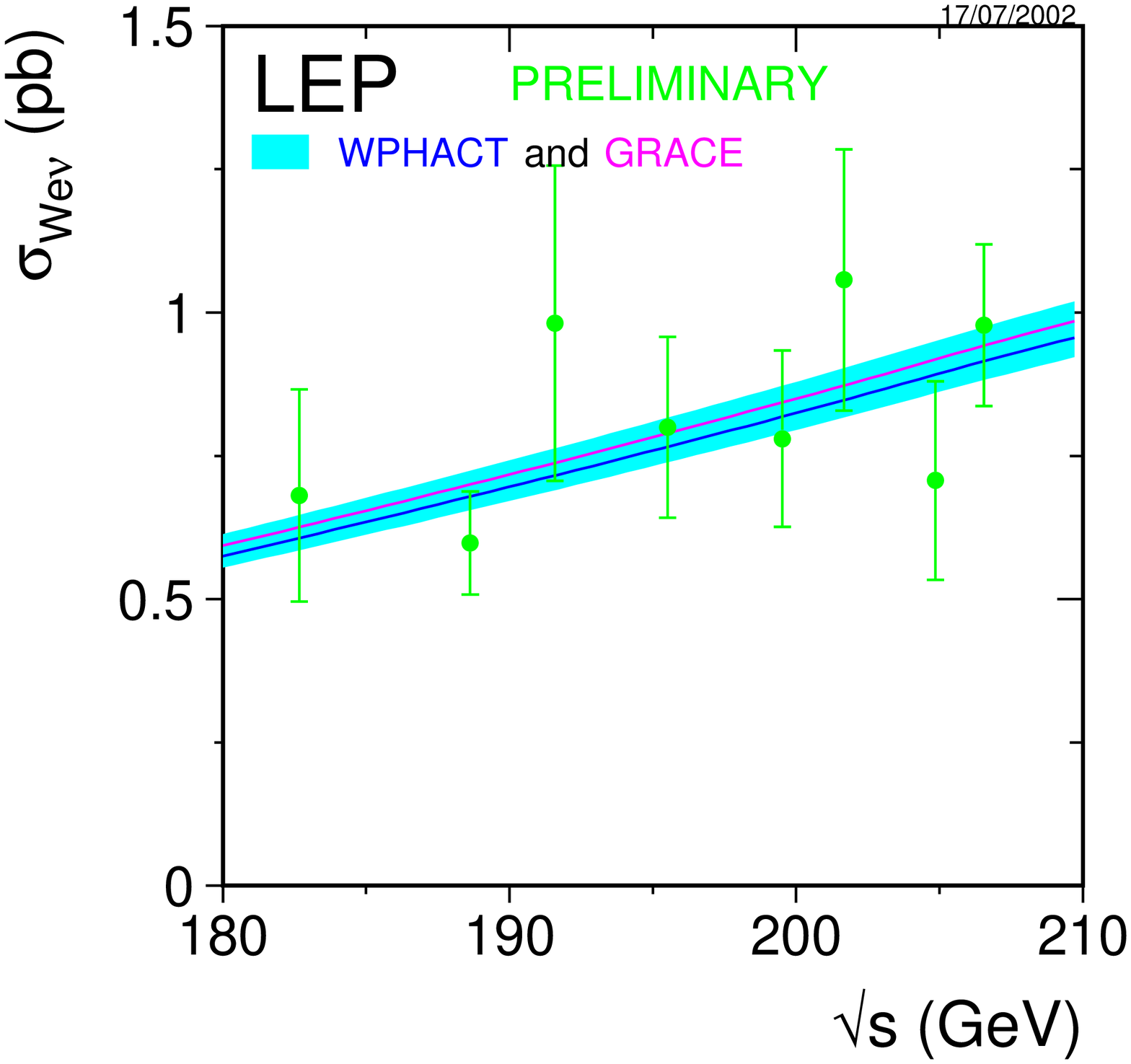,height=8cm}
              \epsfig{file=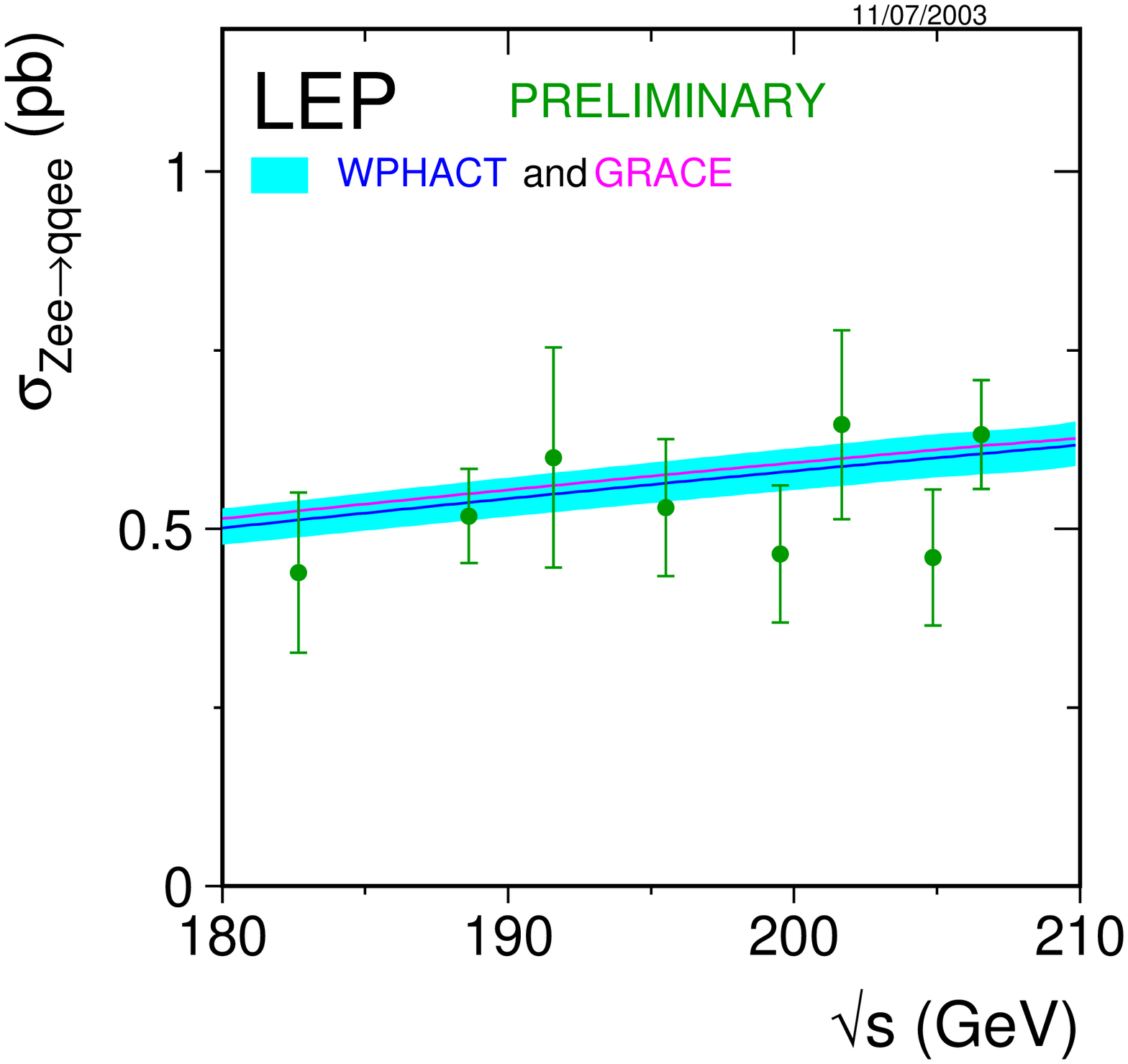,height=8cm}}
\caption{\sl Single W and single Z cross-sections measured at LEP 
at $\rs$=161-207~GeV, compared to SM predictions.}
\label{fig:xsec2}
\end{figure}

Four-fermion final states arising from single W and Z productions
have also been measured in the LEP2 data~\cite{ew}, and are
shown in Fig.~\ref{fig:xsec2}.
Results are in agreement with SM expectations within the
combined experimental precision of 7-8\%.
  
\section{Constraints on Gauge Self Couplings}
The structure of charged TGC vertices WWZ and WW$\gamma$ 
have been studied using W-pair, single W 
and single $\gamma$ events,
and possible deviations from SM expectations have been searched for.
The most common set of parameters to describe the two vertices is
$g_1^{\rm Z}$, $\lambda_\gamma$ and $\kappa_\gamma$,
which conserve both C and P, and obey the SM gauge structure.
The combined LEP2 data yields
$g_1^{\rm Z}=0.991\pm0.022$, $\lambda_\gamma=-0.016\pm 0.022$ 
and $\kappa_\gamma=0.984\pm 0.045$, therefore 
in agreement with the SM expectations 
 $\lambda_\gamma=0$ and $\kappa_\gamma=g_1^{\rm Z}=1$ at the 2-5\% level~\cite{ew}.

Charged QGC as WWZZ, WWZ$\gamma$ and WW$\gamma\gamma$
are predicted in the SM but are beyond the sensitivity of the LEP2 data.
Larger anomalous contributions have been searched for in 
WW$\gamma$ and $\nu\nu\gamma\gamma$ final states, leading to limits on 
the new physics scale of the anomalous contributions 
$\Lambda>$5-10~GeV at 95\% confidence level~\cite{ew}.

Neutral TGC as ZZZ, ZZ$\gamma$ and Z$\gamma\gamma$ do not 
exist in the SM but possible contributions have been searched for 
in Z pair, Zee and Z$\gamma$ events.
No evidence for neutral TGC has been found and parameterizing the vertices
with the $h_i^V$ and $f_i^V$ ($i=1,4$ and $V=\gamma$, Z) couplings,  
limits have been set at the level  $h_i^V, f_i^V<$0.05-0.20
at 95\% confidence level~\cite{ew}.

Neutral QGC also do not exist in the SM, but possible contributions
of a ZZ$\gamma\gamma$ vertex have been searched for in the LEP2 data.
Here the study of ZZ$\gamma$ and $\nu\nu\gamma\gamma$ events
allows to set limits on
the new physics scale of the anomalous contributions
again with $\Lambda>$5-10~GeV  at 95\% confidence level~\cite{ew}.




%

\end{document}